\documentclass{IEEEtran}
\usepackage{amsmath,amssymb,amsthm,cite,multirow,verbatim,url,graphicx}
\usepackage[algo2e]{algorithm2e}
\fussy
\newtheorem{lemma}{Lemma}

\theoremstyle{definition}
\newtheorem{algorithm}{Algorithm}

\providecommand{\GF}{\mathrm{GF}}

\title{Cyclotomic FFTs with Reduced Additive Complexities Based on a Novel Common Subexpression Elimination Algorithm}
\author{
Ning~Chen,~\IEEEmembership{Student Member,~IEEE} and Zhiyuan~Yan,~\IEEEmembership{Senior Member,~IEEE}
\thanks{This work was supported in part by Thales Communications Inc. and in part by a grant from the Commonwealth of Pennsylvania, Department of Community and Economic Development, through the Pennsylvania Infrastructure Technology Alliance~(PITA).
The material in this paper was presented in part at the IEEE Workshop on Signal Processing Systems, Shanghai, China, October 2007.}
\thanks{The authors are with the Department of Electrical and Computer Engineering, Lehigh University, Bethlehem, PA 18015, USA (email: nic6@lehigh.edu; yan@lehigh.edu).}
}
\begin{document}


\maketitle
\begin{abstract}
In this paper, we first propose a novel common subexpression elimination~(CSE) algorithm for matrix-vector multiplications over characteristic-2 fields. As opposed to previously proposed CSE algorithms, which usually focus on complexity savings due to recurrences of subexpressions, our CSE algorithm achieves two types of complexity reductions, differential savings and recurrence savings, by taking advantage of the cancelation property of characteristic-2 fields. Using our CSE algorithm, we reduce the additive complexities of cyclotomic fast Fourier transforms~(CFFTs). Using a weighted sum of the numbers of multiplications and additions as a metric, our CFFTs achieve smaller total complexities than previously proposed CFFTs and other FFTs, requiring both fewer multiplications and fewer additions in many cases.
	\end{abstract}
\section{Introduction}
Discrete Fourier transforms~(DFTs) over finite fields have widespread applications in error correction coding \cite{Blahut79}.
For Reed--Solomon codes, all syndrome-based bounded distance decoding methods involve DFTs over finite fields \cite{Blahut79}: syndrome computation and the Chien search are both evaluations of polynomials and hence can be viewed as DFTs;
inverse DFTs are used to recover transmitted codewords in transform-domain decoders.
Thus efficient DFT algorithms can be used to reduce the complexity of Reed--Solomon decoders.
For example, using the prime-factor fast Fourier transform~(FFT) in \cite{Truong06}, Truong {\it et al.} proposed \cite{Truong06a} an inverse-free transform-domain Reed--Solomon decoder with substantially lower complexity than time-domain decoders;
FFT techniques are used to compute syndromes for time-domain decoders in \cite{Lin07}.

Using an approach similar to those in previous works (see, for
example, \cite{Zakharova92}), cyclotomic FFT~(CFFT) was recently
proposed \cite{Trifonov03} and two variants were subsequently
considered \cite{Costa04,Fedorenko06}. To avoid confusion, we refer
to the CFFT proposed in \cite{Trifonov03} as direct CFFT~(DCFFT) and those in \cite{Costa04} and \cite{Fedorenko06} as
inverse CFFT~(ICFFT) and symmetric CFFT~(SCFFT) respectively henceforth in this paper.
DCFFT has been shown to be efficient for full DFTs of lengths up to 511
\cite{Trifonov03}, and ICFFT and SCFFT are particularly suitable
for \emph{partial DFTs}, which compute only \textbf{part} of the
spectral components and are important for such operations as
syndrome computation of Reed--Solomon decoders \cite{Costa04,Fedorenko06}.

Although CFFTs in \cite{Trifonov03,Costa04,Fedorenko06}
achieve low multiplicative complexities, their additive complexities
(numbers of additions required) are very high if implemented
directly.
The methods used in \cite{Trifonov03,Costa04,Fedorenko06} somewhat alleviate the problem, but the additive complexities of CFFTs in \cite{Trifonov03,Costa04,Fedorenko06} remain quite high.
In this paper, we first propose a novel common subexpression
elimination~(CSE) algorithm, and then use it to reduce the additive
complexities of various CFFTs. The contributions of this paper
are:
\begin{itemize}
	\item To minimize the additive complexities of
CFFTs is a special case of the well-known collection-of-sums
problem, which is NP-complete \cite{Garey79,Cappello84}.
Aiming to reduce additive complexities, previously proposed CSE
algorithms focus primarily on identifying recurring subsets of
summands (we refer to this as subexpressions or patterns). In
contrast, our CSE algorithm, which has only polynomial complexity,
also takes advantage of two other types of complexity reductions
enabled by the underlying characteristic-2 fields: in addition to
explicit recurring subexpressions mentioned above, our CSE algorithm
also considers implicit subexpressions for additional savings;
since the difference between two sums may require fewer additions than one of the two sums, our CSE algorithm also captures savings of this type.
\item We investigate the properties of the three types of CFFTs mentioned above and establish the relations among them.
We first show that the three types of CFFTs have the same
multiplicative complexities assuming the same bilinear forms.
Furthermore, we establish that, \textbf{under direct implementation}, all three types of CFFTs have the same additive complexities.
Finally, we show that there is a mapping between SCFFTs and ICFFTs that preserves the additive complexities \textbf{regardless of implementation}.
Thus, from the perspective of both multiplicative and additive complexities, SCFFTs and ICFFTs are equivalent.
Our results simplify the analysis of their multiplicative and additive complexities as well as performance comparison.
\item Using our CSE algorithm, we reduce the additive complexities of full CFFTs greatly.
In comparison to the full CFFTs in \cite{Trifonov03,Costa04,Fedorenko06}, the best results to our knowledge, our CFFTs have 4\%--15\% smaller additive complexities while maintaining the same multiplicative complexities.
Compared to some previously proposed FFTs techniques, our CFFTs require fewer multiplications \textbf{and} fewer additions.
In comparison to some other FFTs techniques, our CFFTs require fewer multiplications but more additions;
in such cases, the total complexities, obtained by assuming that a multiplication over $\GF(2^m)$ is as complex as $2m-1$ additions, of our CFFTs are smaller.
\end{itemize}

The rest of the paper is organized as follows.
In Section~\ref{sec:background}, we briefly review various CFFTs and CSE algorithms to make this paper self-contained.
Section \ref{sec:alg} presents our CSE algorithm.
We investigate the properties of and relations among the three types of CFFTs in
Section~\ref{sec:var}.
CFFTs with reduced additive complexities are obtained by using our CSE algorithm and presented in Section~\ref{sec:fft}.
\section{Background}\label{sec:background}
\subsection{Cyclotomic FFTs}\label{sec:rev}
Given a primitive element
$\alpha \in \mathrm{GF}(2^m)$, the DFT of a vector $\boldsymbol{f} =
(f_0,f_1,\dotsc,f_{n-1})^T$ is defined as
$\boldsymbol{F}\triangleq\bigl(f(\alpha^0), f(\alpha^1), \dotsc,
\ifCLASSOPTIONtwocolumn
\\
\fi
f(\alpha^{n-1})\bigr)^T$, where $f(x) \triangleq \sum_{i=0}^{n-1}f_i x^i
\in \mathrm{GF}(2^m)[x]$\footnote{In this paper, vectors and matrices are represented by boldface letters, and scalars by normal letters.}.
A new cyclotomic FFTs algorithm was
proposed in \cite{Trifonov03}, and for short lengths (up to 511 \cite{Trifonov03}) it is computationally efficient.
Representing $f(x)$ as a sum of linearized polynomials by cyclotomic
decomposition \cite{Trifonov03,Fedorenko06}, cyclotomic FFT
$\boldsymbol{F} = \boldsymbol{AL f}'= \boldsymbol{AL\Pi f}$, where
$\boldsymbol{A}$ is an $n\times n$ binary matrix,
$\boldsymbol{L}=\mathrm{diag}(\boldsymbol{L}_0, \boldsymbol{L}_1,
\dotsc,\boldsymbol{L}_{l-1})$ is a block diagonal matrix with square
matrices $\boldsymbol{L}_i$'s on the diagonal, $l$ is the number of
cyclotomic cosets, $\boldsymbol{f}'=(\boldsymbol{f}'^T_0,
\boldsymbol{f}'^T_1, \dotsc, \boldsymbol{f}'^T_{l-1})^T$ is a
permutation of the input vector $\boldsymbol{f}$, and
$\boldsymbol{\Pi}$ is a permutation matrix. Suppose
$\boldsymbol{L}_i$ corresponds to a coset of size $m_i$, using a
normal basis of $\GF(2^{m_i})$ generated by $\gamma_i$, then
$\boldsymbol{L}_i$ becomes a circulant matrix \cite{Horn85}:
\begin{equation}
		\boldsymbol{L}_i = \begin{bmatrix}
				\gamma_i^{2^0} & \gamma_i^{2^1} & \dotso & \gamma_i^{2^{m_i-1}}\\
				\gamma_i^{2^1} & \gamma_i^{2^2} & \dotso & \gamma_i^{2^0}\\
				\vdots & \vdots & \ddots & \vdots\\
				\gamma_i^{2^{m_i-1}} & \gamma_i^{2^0} & \dotso & \gamma_i^{2^{2m_i-2}}
		\end{bmatrix}.\label{eqn:circulant}
\end{equation}
Henceforth in this paper we assume $\boldsymbol{L}_i$'s in
$\boldsymbol{L}$ are always constructed by normal bases and we say
$\boldsymbol{L}_i$ in \eqref{eqn:circulant} is a circulant matrix
generated by $\gamma_i$. Thus the product of $\boldsymbol{L}_i$ and
$\boldsymbol{f}'_i$ can be computed as a cyclic convolution, for
which fast bilinear form algorithms are available
\cite{Winograd77,Wagh83,Blahut83,Blahut84}. These fast algorithms
can be written in matrix form as $\boldsymbol{L}_i
\boldsymbol{f}'_i=\boldsymbol{Q}_i(\boldsymbol{R}_i\boldsymbol{b}_i
\cdot\boldsymbol{P}_i\boldsymbol{f}'_i)=\boldsymbol{Q}_i
(\boldsymbol{c}_i\cdot\boldsymbol{P}_i\boldsymbol{f}'_i)$, where
$\boldsymbol{b}_i=(\gamma_i, \gamma_i^2, \dotsc,
\gamma_i^{2^{m_i-1}})^T$, $\boldsymbol{Q}_i$, $\boldsymbol{R}_i$,
and $\boldsymbol{P}_i$ are binary matrices,
$\boldsymbol{c}_i=\boldsymbol{R}_i\boldsymbol{b}_i$ is a
precomputed constant vector, and $\cdot$ stands for pointwise
multiplications. Combining all the terms, a DCFFT is given by
$\boldsymbol{F} = \boldsymbol{AQ}(\boldsymbol{c}\cdot\boldsymbol{P
f}')$, where $\boldsymbol{Q}$ and $\boldsymbol{P}$ are both block
matrices, for which the blocks off the diagonal are the zero
matrices and the diagonal blocks are $\boldsymbol{Q}_i$'s and
$\boldsymbol{P}_i$'s respectively, and
$\boldsymbol{c}=(\boldsymbol{c}_0^T, \boldsymbol{c}_1^T, \dotsc,
\boldsymbol{c}_{l-1}^T)^T$. We remark that both $\boldsymbol{Q}$ and
$\boldsymbol{P}$ are binary and usually sparse. For details of
CFFTs, please refer to \cite{Trifonov03}.

Two variants of CFFTs were proposed in \cite{Costa04,Fedorenko06}.
First, by using the same permutation for both $\boldsymbol{F}$ and
$\boldsymbol{f}$, SCFFTs proposed in \cite{Fedorenko06} satisfy
$\boldsymbol{F}' =
\boldsymbol{L}^T\boldsymbol{A}'^T\boldsymbol{f}'$, where
$\boldsymbol{F}'=\boldsymbol{\Pi F}$ and
$\boldsymbol{f}'=\boldsymbol{\Pi f}$. SCFFTs are so named because
they have symmetric transform matrices, that is, $
\boldsymbol{L}^T\boldsymbol{A}'^T=\boldsymbol{A}'\boldsymbol{L}$. It
is easy to deduce that $\boldsymbol{A}'=\boldsymbol{\Pi A}$. ICFFTs,
proposed in \cite{Costa04}, are based on inverse DFTs and satisfy
$\boldsymbol{F}'' =
\boldsymbol{L}^{-1}\boldsymbol{A}^{-1}\boldsymbol{f}$, where
$\boldsymbol{F}''$
is also
a permutation of $\boldsymbol{F}$. Both SCFFTs and ICFFTs require
fewer multiplications than DCFFTs for partial DFTs, where only a
subset of components in $\boldsymbol{F}$ are needed.

The multiplicative complexity of each CFFT, i.e., the number of
multiplications required, is the total number of non-trivial scalar
multiplications in all cyclic convolutions. That is, the
multiplicative complexity of $\boldsymbol{c}_i\cdot
\boldsymbol{P}_i\boldsymbol{f}'_i$ is the number of non-one elements
in $\boldsymbol{c}_i$ (no element is zero in $\boldsymbol{c}_i$),
which is determined by the cyclic convolution algorithms.
To find the optimum cyclic convolution algorithms with the minimum
multiplicative complexities in CFFTs is still an open problem. In this paper,
we use the cyclic convolution algorithms in \cite{Trifonov}.

The additive complexity of each CFFT is determined by the two
matrix-vector multiplications in which both matrices are binary. For
example, in DCFFTs, the matrices are $\boldsymbol{AQ}$ and
$\boldsymbol{P}$. Due to the large size of $\boldsymbol{AQ}$, direct
computation of the matrix-vector product will result in high
additive complexity. A heuristic algorithm based on erasure decoding
\cite{Trifonov07} was used in \cite{Trifonov03} to reduce the
additive complexity. Similar optimization was also used in
\cite{Costa04}. Another fast matrix-vector multiplication algorithm
is the Four Russians' algorithm \cite{Aho74}, but it is based on
preprocessing and fails to efficiently exploit the matrix structure.
CSE is another commonly used technique for fast
matrix-vector multiplication.

\subsection{Common Subexpression Elimination}\label{sec:cse}
Consider a linear transform $\boldsymbol{Y}=\boldsymbol{MX}$, where
$\boldsymbol{Y}$ and $\boldsymbol{X}$ are $n$- and $n'$-dimensional
column vectors, respectively, and $\boldsymbol{M}$ is an $n \times n'$ matrix
containing only 1, -1, and 0.
Clearly, such a transform requires only additions and subtractions.
It was shown that it is an NP-complete problem \cite[Ensemble Computation]{Garey79}, \cite[Collection of Sums]{Cappello84} to minimize the number of additions and subtractions.

A special type of the collection-of-sums problem is the MCM problem \cite{Potkonjak96}, where the relative position of a bit pattern within the matrix is of no importance \cite{Pasko99}.
This is a valid assumption in the case of the $\boldsymbol{X} = (c^0 x_0, c^1 x_0, \dotsc, c^{n-1} x_0)^T$ with $c=2$ or $c = 2^{-1}$, which is common in filters.
Thus, patterns that differ in relative positions only can be obtained from one of them by shift operations.
This class of problems have wide applications in finite impulse response (FIR) filters~\cite{Potkonjak96,Pasko99,Hartley96,Martinez-Peiro02,Mahesh08,Dempster95,Chang08}.
Graph-based algorithms \cite{Dempster95,Chang08} synthesize directed acyclic graphs, in which partial sums define nodes and shifts are annotated on edges.
In~\cite{Dempster95}, optimal solutions can be obtained by exhaustive search of all topologies with high computational complexity.
Entropy and conditional entropy are used in \cite{Chang08} for vertex decomposition.
Pattern-based algorithms \cite{Potkonjak96,Pasko99,Zhang02,Hartley96,Martinez-Peiro02,Mahesh08,Macleod04} reduce the MCM complexity by first identifying recurring patterns, which are combinations of non-zero positions, and then calculating them only once. They usually use canonical signed digit (CSD) representation to identify potentially sharable subexpressions.
The algorithms in \cite{Potkonjak96,Hartley96,Martinez-Peiro02,Zhang02,Macleod04,Mahesh08} use weight-2 subexpressions as the primitive elements. In contrast, \cite{Pasko99} searches for the highest-weight common subexpressions.

To minimize the number of additions in the matrix-vector multiplications in CFFTs over characteristic-2 fields constitutes a different type of the collection-of-sums problem, where $\boldsymbol{X}$ is over characteristic-2 fields. This implies two properties: (1) the summands are independent, and (2) the $1$'s in $\boldsymbol{M}$ are equivalent to $-1$ and additions are equivalent to subtractions. The second property is noted but not utilized in \cite{Potkonjak96}. Due to the two properties, CSE techniques for the MCM problem are not suitable for the problem considered in this paper.  Heuristic CSE techniques proposed in the context of MCM problem result in modest complexity reductions since they do not take advantage of the second property (see, for example, \cite{Potkonjak96}). For the first property, sophisticated CSE techniques, especially those relying on CSD representation,  that are tailored for FIR filters (see, for example, \cite{Pasko99,Hartley96,Martinez-Peiro02,Mahesh08,Dempster95,Chang08}) \textbf{cannot be directly applied} to our problem; to adapt these algorithms to our problem is not straightforward and requires nontrivial research efforts. Hence, we do not consider CSE techniques for the MCM problem in this paper.

Two CSE algorithms that account for the cancelation property were proposed in \cite{Trifonov03,Gustafsson07}, and we will compare our CSE algorithms against these.

\section{A Novel CSE Algorithm over Characteristic-2 Fields}\label{sec:alg}
We propose a novel CSE algorithm with polynomial complexity that significantly reduces the additive complexities of CFFTs.
Although our CSE algorithm does not guarantee to minimize the additive complexities, it may do so in some cases, especially when the size of the problem is small.

Let us establish the terminology to describe our CSE algorithms.
For a matrix-vector multiplication $\boldsymbol{Y=MX}$, where $\boldsymbol{Y}=(Y_0, Y_1, \dotsc, Y_{n-1})^T$ and $\boldsymbol{X}=(X_0, X_1, \dotsc, X_{n'-1})^T$ are $n$- and $n'$-dimensional column vectors and $\boldsymbol{M}$ is an $n \times n'$ matrix, we refer to the components in $\boldsymbol{Y}$ as sums and the components in $\boldsymbol{X}$ as summands.
Note that the sums in $\boldsymbol{Y}$ have one-to-one correspondence with the rows in $\boldsymbol{M}$, and in direct computation the number of additions required to compute a sum is the number of ones in its corresponding row minus one.
Hence, with a slight abuse of
terminology, we sometimes use rows and sums in an exchangeable
manner. Similarly, there is a one-to-one correspondence between the
summands and the columns in $\boldsymbol{M}$, and we sometimes use
columns and summands in an exchangeable fashion below.

Our CSE algorithm achieves two kinds of savings: differential
savings and recurrence savings, as defined in Sections
\ref{sec:diff} and \ref{sec:recur}, respectively.

\subsection{Differential Savings}\label{sec:diff}
Let $\boldsymbol{Y=MX}$ represent a matrix-vector multiplication,
in which $\boldsymbol{X}$ is over characteristic-2 fields and
$\boldsymbol{M}$ is a binary matrix. For the column positions where
$\boldsymbol{M}_{r_p}$ and $\boldsymbol{M}_{r_c}$, rows $r_p$ and
$r_c$ ($r_p\neq r_c$) of $\boldsymbol{M}$ respectively, both have
ones, the difference (or sum)
$\boldsymbol{M}_{r_c}-\boldsymbol{M}_{r_p}$ of the two rows has
zeros. If $\boldsymbol{M}_{r_c}-\boldsymbol{M}_{r_p}$ contains fewer
entries than one of the two rows, say $\boldsymbol{M}_{r_c}$, we can
reduce the total number of additions by first computing $Y_{r_p}$
and then computing
$Y_{r_c}=Y_{r_p}+(\boldsymbol{M}_{r_c}-\boldsymbol{M}_{r_p})\boldsymbol{X}$.
Let us denote the numbers of non-zero entries in
$\boldsymbol{M}_{r_p}$, $\boldsymbol{M}_{r_c}$, and
$\boldsymbol{M}_{r_c}-\boldsymbol{M}_{r_p}$ as $w_p$, $w_c$, and
$w_d$, respectively, the differential saving (the number of
additions saved) is given by $w_c-w_d-1$. Since we are only
concerned about positive savings, we use $(w_c-w_d-1)^+ \triangleq
\max\{0, w_c-w_d-1\}$ in our algorithms.

The price for the differential saving is that now $Y_{r_p}$ must be computed before $Y_{r_c}$, putting a dependency between the two sums.
We use an ordered pair $(r_p, r_c)$ to represent this dependency;
we call $Y_{r_p}$, the sum computed first, the parent, and refer to $Y_{r_c}$ as the child.
Since each ordered pair introduces a dependency, to keep track of all dependency, we use a digraph to keep track of all ordered pairs, where the vertices are the row numbers in the ordered pairs and the edges are from the parent to the child in each pair.
We call this graph \emph{dependency graph} henceforth in this paper.
There is no conflicting dependency as long as the dependency graph is acyclic.
Thus, before any ordered pair can be added to the dependency graph, it is necessary to check whether the addition of the new ordered pair will introduce cycles in the dependency graph;
if yes, this ordered pair is called cycle-inducing and hence not permissible.
Cycle detection can be done recursively.

When an ordered pair $(r_p, r_c)$ is added to the dependency graph,
both $\boldsymbol{M}$ and $\boldsymbol{X}$ need to be transformed.
We first append $Y_{r_p}$ to $\boldsymbol{X}$ as a new summand. We
also replace $\boldsymbol{M}_{r_c}$ with the difference
$\boldsymbol{M}_{r_c}-\boldsymbol{M}_{r_p}$;
then due to the new
summand $Y_{r_p}$, a new column with a single one at the $r_c$th
position and zeros at other positions is appended to
$\boldsymbol{M}$. We call these operations a differential
transformation.

Our differential transformations bear some similarities to the
erasure correction approach used in \cite{Trifonov07}. As pointed
out in \cite{Trifonov07},
$\boldsymbol{Y}=\boldsymbol{M}\boldsymbol{X}$ is equivalent to
$[\boldsymbol{M} \mid \boldsymbol{I}](\boldsymbol{X}^T,
\boldsymbol{Y}^T)^T=\boldsymbol{0}$, which defines a code
$\mathcal{C}$ with all codewords $(\boldsymbol{X}^T,
\boldsymbol{Y}^T)^T$;
to compute $\boldsymbol{Y}=\boldsymbol{M}\boldsymbol{X}$ is equivalent to
erasure correction with $\boldsymbol{Y}$ erased based on
$\mathcal{C}$. After a series of differential transformations as
described above, the matrix-vector multiplication becomes
$\boldsymbol{Y}=\boldsymbol{M}'\boldsymbol{X}'$, where
$\boldsymbol{X}'=(\boldsymbol{X}^T, \boldsymbol{Y}'^T)^T$,
$\boldsymbol{Y}'$ consists of the summands corresponding to all the
parents in the ordered pairs, and $\boldsymbol{M}'$ has the same
number of rows as $\boldsymbol{M}$. By adding all-zero columns to
$\boldsymbol{M}'$, we can find a matrix $\boldsymbol{M}''$ such that
$\boldsymbol{Y}=\boldsymbol{M}''(\boldsymbol{X}^T,
\boldsymbol{Y}^T)^T$. Hence $(\boldsymbol{M}'' - [\boldsymbol{0}
\mid \boldsymbol{I}])(\boldsymbol{X}^T, \boldsymbol{Y}^T)^T =
\boldsymbol{0}$. Thus our differential transformations lead to a
different parity check matrix
for the same code
$\mathcal{C}$. Furthermore, the acyclic property for the dependency
graph ensures that $\boldsymbol{Y}$ can be recovered by using the
parity check matrix $\boldsymbol{M}'' - [\boldsymbol{0} \mid
\boldsymbol{I}]$. From this perspective our differential
transformation is similar to that of the message passing part of
\cite{Trifonov07}: both find an alternative parity check matrix with
smaller Hamming weights for the code $\mathcal{C}$, which can be
used to compute $\boldsymbol{Y}$. However, different search methods
are used to obtain alternative parity check matrices in our work and
in \cite{Trifonov07}.
\subsection{Recurrence Savings}\label{sec:recur}
We refer to the number of occurrences of a subexpression (or pattern
in the rows of $\boldsymbol{M}$) as \emph{pattern frequency}, and
define the \emph{recurrence saving} of each pattern as its pattern
frequency minus $1$. After a subexpression is identified, we append
the subexpression to $\boldsymbol{X}$ as a new summand, and
$\boldsymbol{M}$ is updated accordingly. These operations are
referred to as a recurrence transformation. A sequence of recurrence
transformations can be described in a matrix decomposition form: $\boldsymbol{M}=\boldsymbol{M}_R\prod_{i=0}^{K-1}\boldsymbol{T}_i,$
where $\boldsymbol{T}_i = [\boldsymbol{I}\mid\boldsymbol{G}_i^T]^T$,
the row vector $\boldsymbol{G}_i$ corresponds to a subexpression,
$\boldsymbol{M}_R$ has no pattern recurrence, and $K$ is the number
of identified subexpressions. Thus $\boldsymbol{Y}$ is computed in a
sequential fashion: first assign $\boldsymbol{X}^{(0)} =
\boldsymbol{X}$, then compute $\boldsymbol{X}^{(i+1)}=
\boldsymbol{T}_i \boldsymbol{X}^{(i)}$ for $i = 0, 1, \dotsc, K-1$,
and finally compute $\boldsymbol{Y}=\boldsymbol{M}_R
\boldsymbol{X}^{(K)}$. For $0\leq i \leq K-1$, let
$\boldsymbol{M}^{(i)}$ denote $\boldsymbol{M}_R
\prod_{l=i}^{K-1}\boldsymbol{T}_l$, and
$\boldsymbol{Y}=\boldsymbol{M}^{(i)}\boldsymbol{X}^{(i)}$.

Compared with the matrix splitting method \cite{Pasko99}, recurrence
transformations keep track of the identified subexpressions as new
summands, instead of simply removing them. To reduce the
computational complexity of pattern search, our CSE algorithm looks
for only two-summand subexpressions. However, since each two-summand
subexpression is in turn appended as a summand and multi-summand
subexpressions can be expressed recursively as two-summand
subexpressions, our CSE algorithm efficiently exploits the
recurrence savings of both two-summand patterns and multi-summand
patterns.

One limitation of the recurrence transformations above is
that it considers only explicit subexpressions,
missing implicit subexpressions that are hidden by cancelation.
We will now identify implicit subexpressions through \emph{forced
patterns}. To this end, after a two-summand pattern $X_0+ X_1$ is
identified and introduced as a new summand $X_n$, we try to impose
the pattern on the rows containing \textbf{only} $X_0$ or $X_1$ by
replacing $X_0$ with $X_1+X_n$ or $X_1$ with $X_0+X_n$. After
forcing patterns $X_1+X_n$ or $X_0+X_n$ on row $r_i$, if previously
identified patterns emerge due to cancelation and therefore lead to
complexity savings, we transform $\boldsymbol{M}_{r_i}$ to reflect
the forced pattern. If the forced pattern does not lead to any
saving, we do not transform $\boldsymbol{M}_{r_i}$. Since a forced
pattern leads to complexity saving \textbf{only when} they match
previously identified patterns, we search the rows only for
previously identified patterns. Since we keep track of all
two-summand patterns, we first search the rows for previously
identified patterns that include $X_0$ or $X_1$, which is inserted
due to the forced pattern. If we find a previously identified
pattern, say $X_j=X_i+X_0$, in row $r_i$, we replace $X_i+X_0$ by
$X_j$ and continue to search for all previously identified patterns
that include $X_j$, and so on.

Now we illustrate the advantage of the forced pattern method by a simple
example. Say we have established three patterns as $X_4 =X_1+X_2$,
$X_5=X_3+X_4$, and $X_6=X_0+X_1$. Now let us consider the sum $Y_0 =
X_0+ X_2+X_3$, which does not contain the identified patterns $X_4$, $X_5$, or $X_6$
explicitly. But if we force $X_6$ on $Y_0$, we have
$Y_0=X_1+X_2+X_3+X_6$, which becomes $Y_0=X_5+X_6$ after replacing
 previously identified patterns $X_1+X_2$ with $X_4$ and $X_3+X_4$ with $X_5$
as described above. In this simple example, by forcing the pattern
we reduce the number of additions by one. In a nutshell, it is a
greedy strategy in which, based on existing subexpressions, we try
to find an alternative expression that requires fewer additions for
a sum.

When introducing forced patterns for a sum, new summands for the sum
are introduced. If any new summand is a sum, this introduces
dependency between the two sums, and possibly cycles in the
dependency graph. We replace $X_1$ with $Y_1$ in the simple example
above to illustrate such a case.
If we force the pattern $X_6=X_0+Y_1$ pattern on $Y_0$, we have $Y_0=Y_1+X_2+X_3+X_6=X_5+X_6$. Although it reduces the number of additions by one, it requires that $Y_1$ should be computed before $Y_0$.
Since forced patterns introduce new dependency, we will keep track
of this using the dependency graph and cycle detection
is necessary in recurrence transformations if we consider forced patterns.

\subsection{Approximate Dynamic Programming}\label{sec:ralg}
We have discussed two kinds of transformations that result in
differential savings and recurrence savings. A remaining question is:
how should we coordinate the transformations associated with
differential savings and recurrence savings? That is, which
kind of saving is more preferable? A seemingly straightforward
answer would be to use a simple greedy strategy: choose one
transformation with the greatest saving. Instead of this simple
greedy strategy, we adopt a different strategy. We justify our
choice by approximate dynamic programming \cite{Bertsekas95} below.

Note that both differential and recurrence transformations can be
expressed in a matrix decomposition form. Thus the
collection-of-sums problem can be viewed as a dynamic programming
problem \cite{Bertsekas95}, where the cost to be minimized is the
number of additions and each differential or recurrence
transformation corresponds to one stage.
The total cost is denoted by $A = \sum_{i=0}^{K - 1}g^{(i)} + J_R$ where $g^{(i)} \in \{0, 1\}$ is the cost of Stage $i$ and $J_R$ is the cost of implementing $\boldsymbol{M}_R$.
Let us denote $\boldsymbol{M}$ and $\boldsymbol{X}$ after the $i$th stage as the $\boldsymbol{M}^{(i)}$ and $\boldsymbol{X}^{(i)}$, and they are the state variables.
The idea of approximate dynamic programming is to approximate and
optimize the cost-to-go $J$ \cite{Bertsekas95}. Suppose after the
transformations in Stage $i$, the matrix-vector multiplication is
given by $\boldsymbol{Y}=\boldsymbol{M}^{(i)}\boldsymbol{X}^{(i)}$.
Since under direct computation, it needs $W(\boldsymbol{M}^{(i)})-n$
additions, where $W(\boldsymbol{M}^{(i)})$ is the number of 1's in
$\boldsymbol{M}^{(i)}$, we use $J^{(i)} = a^{(i)}
\bigl(W(\boldsymbol{M}^{(i)})-n\bigr)$ as a linear approximation of the
cost-to-go, where $a^{(i)}$ approximates $\bigl(A - \sum_{j=0}^{i - 1}g^{(j)}\bigr)\bigm/
\bigl(W(\boldsymbol{M}^{(i)})-n\bigr)$.
When $a^{(i)}=\bigl(A - \sum_{j=0}^{i-1}g^{(j)}\bigr)\bigm/
\bigl(W(\boldsymbol{M}^{(i)})-n\bigr)$, $J^{(i)}=A - \sum_{j=0}^{i-1}g^{(j)}$ is indeed the cost-to-go.
Suppose for Stage $i$, the largest differential and recurrence savings are $s_d^{(i)}$ and $s_r^{(i)}$, respectively.
Based on the above approximation, we can find a transformation that minimizes the cost-to-go.
If a differential transformation is chosen, the matrix weight after the transformation is given by $W(\boldsymbol{M}^{(i+1)})= W(\boldsymbol{M}^{(i)}) - s_d^{(i)}$;
otherwise, it is $W(\boldsymbol{M}^{(i+1)})=W(\boldsymbol{M}^{(i)}) - s_r^{(i)} - 1$.
Then the approximate optimal cost-to-go is the smaller between $a^{(i)}\cdot \bigl(W(\boldsymbol{M}^{(i)}) - s_d^{(i)} - n\bigr)$ and $1+a^{(i)}\cdot\bigl(W(\boldsymbol{M}^{(i)})-s_r^{(i)}-1-n\bigr)$.
Thus a differential transformation is preferred when $s_d^{(i)} >
s_r^{(i)} + 1 - 1 / a^{(i)}$, and a recurrence transformation is
preferred if $s_d^{(i)} \leq s_r^{(i)} + 1 - 1 / a^{(i)}$. Although
it is difficult to compute $a^{(i)}$ since $A$ is actually unknown,
fortunately the choice between differential saving and recurrence
saving does not require the precise value of $A$. It is obvious that
$0<a^{(i)}<1$ for any $i$, and hence differential transformations
are usually preferred over recurrence transformations even when
$s_d^{(i)} =s_r^{(i)}$. This is particularly the case when $a^{(i)}$
is small. For example, the ratio of the required number of additions
after applying our CSE algorithm,
and $W(\boldsymbol{M}^{(0)})-n$ is between 0.16 and 0.26. Thus,
$a^{(0)}$ is clearly a small fraction. As $i$ increases, $a^{(i)}$
increases while $s_d^{(i)}$ and $s_r^{(i)}$ decrease. Our CSE
algorithm treats the differential transformations with preference in
all cases.

We comment that the simple greedy strategy which selects the greater
one between $s_d^{(i)}$ and $s_r^{(i)}$ mentioned above corresponds
to always setting $a^{(i)}=1$ in approximate dynamic programming,
which does not provide a good approximation. Our simulation results
confirm this observation, as the differential saving first strategy
usually leads to better results than the simple greedy strategy.

Since we are using approximate dynamic programming in every stage,
choosing a differential saving does not take into account all
recurrence savings in future stages. Thus for some stages it may be an unwise
choice. We propose a method to identify such differential savings
and reverse them.
Say $Y_0 = X_0+X_1+X_2+X_3$ and $Y_1 = X_0+X_4+X_5+Y_0$ as a result
of the differential saving from an ordered pair $(0, 1)$. Since $Y_0
+ X_0 = X_1 + X_2 + X_3$, we can replace $X_0+Y_0$ in $Y_1$ by $X_1
+ X_2 + X_3$ and it is clear that $Y_0$ and $Y_1$ have a common
subexpression $X_1 + X_2 + X_3$. Using the subexpression $X_1 + X_2
+ X_3$ effectively reverses the differential transformation
represented by $(0, 1)$. To identify a reversal of this kind, we
search for \emph{reversal patterns};
a reversal pattern $Y_i + X_j$ consists of a sum $Y_i$ and one of its summands $X_j$.
In contrast to other patterns, this pattern may have a recurrence saving of zero, that is, it appears only once.
It can be shown that such a reversal saves only one addition, regardless of the frequency of the reversal pattern;
thus, such a reversal is meaningful only when there are no other subexpressions involving $Y_i$.
For instance, in the above example, if there are more than two recurrences of $Y_0+X_4$, the subexpression $Y_0+X_4$ results in a greater saving than $Y_0 + X_0$.
Thus it will be efficient to search for reversal patterns only after all recurrence savings are accounted for.

Our CSE algorithm, shown below in Algorithm~\ref{alg:cse}, has two major steps, Steps~\ref{alg:cse}.1 and \ref{alg:cse}.3, and they are referred to as the differential and recurrence steps respectively.
\begin{algorithm}Common Subexpression Elimination\label{alg:cse}
	\renewcommand{\labelenumi}{\ref{alg:cse}.\theenumi}
\begin{enumerate}
\item Identify the non-cycle-inducing pairs of rows with the $l_d$ greatest differential savings, select
one pair out of them randomly, and transform both $\boldsymbol{M}$
and $\boldsymbol{X}$ as described above.
\item Repeat Step~\ref{alg:cse}.1 until there is no
differential saving.
\item Identify the two-summand patterns with the $l_r$ greatest recurrence savings, select one
out of them randomly. Replace all occurrences of the selected
pattern with a new entry. On those rows with only one entry of the
pattern, force the pattern if it leads to less 1's in the row.
\item Go to Step~\ref{alg:cse}.1 until there is no recurrence saving.
\item If there is a reversal pattern, reverse the differential saving and go to Step~\ref{alg:cse}.3.
\end{enumerate}
\end{algorithm}
Since differential savings are due to the overlapping ones in two
rows, there is no positive differential saving if there is no
recurrence saving. This is the reason for the termination condition
in Step~\ref{alg:cse}.4. In Steps~\ref{alg:cse}.1 and \ref{alg:cse}.3, we randomly select one transformation
among those with the $l_d$ greatest differential savings and the
$l_r$ greatest recurrence savings respectively. There is a tradeoff
between search space (and hence performance) and search complexity:
greater $l_d$ and $l_r$ enlarge the search space that may lead to greater savings at the expense of higher complexity. In our work, $l_d=l_r=2$ appears
enough for most cases. For matrices with small sizes, the additional
complexity caused by expanding the searching space is usually
affordable. For large matrices, we use $l_d=l_r=1$.
Since Algorithm~\ref{alg:cse} is a randomized algorithm, the result
of each run may vary. However, simulation results show that the
variance between different runs is relatively small in comparison to the number of required additions.

Our sequential transformation of $\boldsymbol{M}$ in Section~\ref{sec:cse}
 is similar to the CSE algorithm in \cite{Zhang02}, but the algorithm in \cite{Zhang02} does not take advantage of the cancelation property of characteristic-2 fields as Algorithm~\ref{alg:cse}.
With forced patterns, Algorithm~\ref{alg:cse} takes advantage of the cancelation property not only by differential savings but also by recurrence savings.
Although Algorithm~\ref{alg:cse} and those in \cite{Trifonov07,Gustafsson07} all take advantage of the cancelation property of characteristic-$2$ fields, they use quite different strategies.
Algorithm~\ref{alg:cse} uses a top-down approach to build the addition sequence by reducing the binary matrix, while a bottom-up approach starting from summands was used in \cite{Gustafsson07}.
The CSE algorithm in \cite{Trifonov07} first rebuilds the binary matrix from low-weight linear combinations of rows, then reduces the matrix top-down using recurrence savings. Also, although the method in \cite{Trifonov07} takes advantage of the cancelation property by erasure decoding in the message passing part, it fails to do so in its CSE part. As we will show in Section~\ref{sec:fft}, Algorithm~\ref{alg:cse} leads to significantly better results than the method in \cite{Trifonov07}.

\subsection{Fast CSE}
When the size of $\boldsymbol{M}$ is large, the time complexity of
Algorithm~\ref{alg:cse} may be prohibitive. We propose several
improvements to reduce the time complexity of
Algorithm~\ref{alg:cse}.

In Algorithm~\ref{alg:cse}, we restart the differential step after
each recurrence step. But the possibility that new differential
savings emerge after we identify a pattern for recurrence saving is
quite small. In order to reduce the complexity, we do not revisit
the differential step after the recurrence step has ended,
essentially decoupling the two steps. This not only reduces the time
complexity by reducing the number of times Step~\ref{alg:cse}.1 is
repeated, but also enables us to further accelerate both steps by
space-time tradeoff, which will be discussed below. Note that our
simulation results show that the decoupling of the two steps results
in only negligible performance loss.

Now that the differential step is standalone, it is necessary to
avoid repeated exhaustive searches. There are only $n$ rows in
$\boldsymbol{M}$, so all possible differential saving can be put in
an $n\times n$ array $\boldsymbol{D}$, where $D_{ij}$ stands for the
differential saving of the ordered pair of rows $(r_i, r_j)$. An
exhaustive search is needed to initialize $\boldsymbol{D}$.
Afterwards, at most $2(n-1)$ entries (namely, the non-diagonal
entries in row $r_i$ and column $r_j$) of the array need to be
updated after each ordered pair $(r_i, r_j)$ is added to the
dependency graph. Whenever one pair of rows is detected to be
cycle-inducing, its differential saving will be set to -1 and hence
it is excluded from future consideration for differential
transformations. As the number of possible pairs decreases
continuously, the search will be increasingly simpler.

A similar idea can be used to reduce the time complexity of the
recurrence step. Since elimination of one pattern will only
change a small portion of the pattern frequencies, to expedite
searches, we store the recurrence savings and update them after each
recurrence transformation.

Because not all patterns exist and the number of possible patterns
will decrease continuously, it will require less storage space if we
keep track of only the patterns with positive recurrence savings.
However, this will involve an exhaustive search to update the
pattern frequencies each time after a pattern is identified, which
may results in high time complexity when the size of
$\boldsymbol{M}$ is large. Instead, we keep track of all pattern
frequencies, including those with no recurrence savings, in a
two-dimensional array $\boldsymbol{R}$. Suppose after the
differential steps are over, and $\boldsymbol{M}'$ has $\bar{n}$
columns. Initially, $\boldsymbol{R}$ is an upper triangle array with
$\bar{n}-1$ rows, where $R_{ij}$ is the recurrence
saving of the two-summand pattern $X_i + X_{i+j+1}$ for $0\leq i
\leq \bar{n}-2$ and $0\leq j \leq \bar{n}-i-2$. The recurrence
saving array $\boldsymbol{R}$ is arranged in this fashion so that
frequency updates can be done by direct addressing without search
and it is not necessary to remove frequencies.
When a new pattern is identified, the two-summand pattern becomes
the $(\bar{n}+1)$th summand. Thus, the frequency of the $i$th and
the $(\bar{n}+1)$th summands is appended to the $i$th row.
Furthermore, a new row with only one element, the frequency of the
$\bar{n}$th and $(\bar{n}+1)$th summands, will be the last row of
$\boldsymbol{R}$. After $X_i+X_j$ is identified as a subexpression,
all frequencies related to $X_i$ or $X_j$ need to be updated. That
is, $R_{i', i - i' - 1}$ for all $i' < i$, $R_{j', j - j' - 1}$ for
all $j' < j$, $R_{i, i'' - i - 1}$ for all $i'' > i$, and $R_{j, j''
- j - 1}$ for all $j'' > j$ are updated accordingly. Furthermore,
$R_{i, j - i - 1}$ is set to zero.

During Step~\ref{alg:cse}.1, our CSE algorithm keeps only one copy
of each row. Actually one row can have different decompositions,
based on differential savings with different rows. To exploit the
best differential saving for each row, a modified differential
saving update scheme is developed.

Let us assume that the ordered pair $(r_p, r_c)$ is selected for
differential transformation, which replaces $\boldsymbol{M}_{r_c}$
with $\boldsymbol{M}'_{r_c}$. For row $r_i$ $(r_i\neq r_p)$, there
are two possible differential savings: one between
$\boldsymbol{M}_{r_c}$ and $\boldsymbol{M}_{r_i}$ and the other
between $\boldsymbol{M}'_{r_c}$ and $\boldsymbol{M}'_{r_i}$. If the
latter is greater, we simply update $D_{ci}$. If the
former is greater, the differential saving $D_{ci}$ is
not changed and $\boldsymbol{M}_{r_c}$ is saved so that it can be
used when $D_{ci}$ is selected.
If two savings are equal, it is
randomly chosen which copy to use. Since we may need different
copies of $\boldsymbol{M}_{r_c}$ for each $r_i$, a three-dimension
array $\boldsymbol{K}$ whose entry $K_{ij}$ keeps a copy of
$\boldsymbol{M}_{r_j}$ corresponding to $D_{ij}$ if
$\boldsymbol{M}_{r_j}$ provides a greater differential saving than
$\boldsymbol{M}'_{r_j}$ with regard to $\boldsymbol{M}_{r_i}$. Since
this can occur recursively, for each row at most $n-1$
different rows may be stored in $\boldsymbol{K}$.

Our CSE algorithm incorporating the above improvements is shown in
Algorithm~\ref{alg:reduction}.
\begin{algorithm}
	Fast CSE\label{alg:reduction}
	\renewcommand{\labelenumi}{\ref{alg:reduction}.\theenumi}
	\begin{enumerate}
		\item Initialize the differential saving array $\boldsymbol{D}$ and $\boldsymbol{K}$.
		\item Find the non-cycle-inducing pairs of rows with the $l_d$ greatest differential savings in $\boldsymbol{D}$, randomly choose one, eliminate it, and update $\boldsymbol{D}$ and $\boldsymbol{K}$ accordingly.
		\item Repeat Step~\ref{alg:reduction}.2 until there is no positive entry in $\boldsymbol{D}$.
		\item Initialize the recurrence saving array $\boldsymbol{R}$.
		\item Find the patterns with the $l_r$ greatest recurrence savings in $\boldsymbol{R}$, randomly choose one, replace all occurrences of it. On those rows with only one entry of the pattern, force the pattern if it leads to less 1's in the row. Update $\boldsymbol{R}$.
	\item Repeat Step~\ref{alg:reduction}.5 until all entries in $\boldsymbol{R}$ are zero.
	\item If there is a reversal pattern, reverse the differential saving, update $\boldsymbol{R}$, and go to Step~\ref{alg:reduction}.5.
	\end{enumerate}
\end{algorithm}

Our simulation results show that after a single run, the difference
between the total additive complexities obtained by
Algorithms~\ref{alg:cse} and \ref{alg:reduction} is negligible.
However, the time complexity of Algorithm~\ref{alg:reduction} is
much smaller than that of Algorithm~\ref{alg:cse}. For example, when
$\boldsymbol{M}$ is a $255 \times 255$ matrix, for a single run
Algorithm~\ref{alg:cse} needs about ten hours while
Algorithm~\ref{alg:reduction} finishes in approximately five minutes.
The difference in run time is greater for matrices with larger
sizes.
Since Algorithms~\ref{alg:cse} and \ref{alg:reduction} are both
probabilistic, the speed advantage of Algorithm~\ref{alg:reduction}
over Algorithm~\ref{alg:cse} enables us to run
Algorithm~\ref{alg:reduction} many more times, enhancing the
possibility of obtaining a better result than using
Algorithm~\ref{alg:cse} within the same amount of time.

\subsection{Example}
Now we provide an example of Algorithm~\ref{alg:reduction}. At the
beginning, $\boldsymbol{K}$ is empty and
\begin{align*}
	\boldsymbol{M} &=
	\begin{bmatrix}
				1 & 0 & 1 & 1 & 1\\
				1 & 1 & 1 & 1 & 1\\
				1 & 1 & 0 & 1 & 1\\
				0 & 1 & 1 & 1 & 0
		\end{bmatrix} &\quad
		\boldsymbol{D} &=
		\begin{bmatrix}
				-1 & 3 & 1 & 0\\
				2 & -1 & 2 & 0\\
				1 & 3 & -1 & 0\\
				0 & 2 & 0 & -1
		\end{bmatrix}.
\end{align*}

Choosing $(0, 1)$ and adding a column corresponding the new summand
$Y_0$, we have
\begin{align*}\boldsymbol{M}^{(1)} &=
		\begin{bmatrix}
				1 & 0 & 1 & 1 & 1 & 0\\
				0 & 1 & 0 & 0 & 0 & 1\\
				1 & 1 & 0 & 1 & 1 & 0\\
				0 & 1 & 1 & 1 & 0 & 0
			\end{bmatrix} &
			\ifCLASSOPTIONonecolumn
			\quad
			\else
			\hskip-0.5em
			\fi
		\boldsymbol{D} &=
	\begin{bmatrix}
				-1 & -1 & 1 & 0\\
				-1 & -1 & 2 & 0\\
				1 & 0 & -1 & 0\\
				0 & 0 & 0 & -1
		\end{bmatrix}.
	\end{align*}
Since $(1, 0)$ is cycle-inducing, its saving is simply set to -1. We
also set $K_{12}$ to $(1, 1, 1, 1, 1)$ to keep track of
$\boldsymbol{M}_1$.

	Choosing $(1, 2)$, the matrices are updated as
	\begin{align*}
		\boldsymbol{M}^{(2)} &=
		\begin{bmatrix}
				1 & 0 & 1 & 1 & 1 & 0 & 0\\
				0 & 1 & 0 & 0 & 0 & 1 & 0\\
				0 & 0 & 1 & 0 & 0 & 0 & 1\\
				0 & 1 & 1 & 1 & 0 & 0 & 0
			\end{bmatrix} &
			\ifCLASSOPTIONonecolumn
			\quad
			\else
			\\
			\fi
			\boldsymbol{D} &=
		\begin{bmatrix}
				-1 & -1 & 0 & 0\\
				-1 & -1 & -1 & 0\\
				-1 & 0 & -1 & 0\\
				0 & 0 & 0 & -1
		\end{bmatrix}.
	\end{align*}

	Note that $(2, 0)$ is cycle-inducing so there is no positive differential saving left.

Now we enter the recurrence transformations. The recurrence saving
array $\boldsymbol{R}$ for $\boldsymbol{M}^{(2)}$ is initialized to
all zeros except that $R_{2, 0}=1$, which corresponds to the pattern
$X_2+X_3$. Hence $\boldsymbol{G}_0$ is $(0, 0, 1, 1, 0, 0, 0)$ and
the algorithm stops at
	\begin{displaymath}
	\boldsymbol{M}_R =
\begin{bmatrix}
				1 & 0 & 0 & 0 & 1 & 0 & 0 & 1\\
				0 & 1 & 0 & 0 & 0 & 1 & 0 & 0\\
				0 & 0 & 1 & 0 & 0 & 0 & 1 & 0\\
				0 & 1 & 0 & 0 & 0 & 0 & 0 & 1
		\end{bmatrix}
	\end{displaymath}
and the recurrence saving array becomes all zeros. The remaining
matrix $\boldsymbol{M}_R$ needs five additions. The identified
pattern $X_2+X_3$ also needs one addition. So
$\boldsymbol{Y}=\boldsymbol{MX}$ can be calculated by six additions,
whereas a straightforward implementation of
$\boldsymbol{Y}=\boldsymbol{MX}$ requires 12 additions. Note that
techniques such as forced patterns or reversal patterns are not
applicable in this simple example.
Nevertheless, since it can be easily verified 
that $\boldsymbol{Y}=\boldsymbol{MX}$ cannot be done in five additions, 
our CSE algorithm minimizes the number of additions in this case.
Note that if we only use recurrence savings, the result will be seven additions.

\subsection{Time and Storage Complexities}
Since the reduction of additive complexities depends on $\boldsymbol{M}$ only, the output of Algorithm~\ref{alg:reduction} for a given CFFT can be used for any input vector.
Hence Algorithm~\ref{alg:reduction} is simply precomputation and its complexity should not be considered as part of the complexities of CFFTs. To show that Algorithm~\ref{alg:reduction} is \textbf{computationally tractable},
we provide an order-of-magnitude analysis for the time and area complexity of Algorithm~\ref{alg:reduction} below.

Algorithm~\ref{alg:reduction} requires only four types of
operations: adding two rows, inserting or removing entries from a
row, searching for a two-summand pattern in a row, and comparison to
find the greatest saving. During the optimization, while the number
of columns in the matrix $\boldsymbol{M}^{(i)}$ increases
continuously, the number of 1's in each row decreases. To facilitate
row additions, for each row we only store the positions of 1's as a
sorted list. Since the original $\boldsymbol{M}$ has $n'$ columns,
adding two rows is equivalent to merging two sorted lists of size at
most $n'$, which requires at most $2n'$ comparisons. For simplicity,
we assume inserting or removing entries in a row has the same
complexity as adding two rows. Searching for a two-summand pattern
in a row needs at most $n'$ comparisons. We assume the complexity of
either appending an entry to a row or updating a matrix entry is
negligible.

Now since differential transformations described in
Steps~\ref{alg:reduction}.1, \ref{alg:reduction}.2, and
\ref{alg:reduction}.3 and recurrence transformations in
Steps~\ref{alg:reduction}.4, \ref{alg:reduction}.5, and
\ref{alg:reduction}.6 are independent, we can analyze them
separately. In Step~\ref{alg:reduction}.1, the initialization of the
differential saving array $\boldsymbol{D}$ needs to add rows for
$n(n-1)/2$ times, so it takes at most $n(n-1)n'$ comparisons. The
result of Step~\ref{alg:reduction}.2 is an acyclic digraph with at
most $n$ nodes, so at most $n(n-1)/2$ pairs of rows are identified.
To identify one pair of rows, we need at most $g-1$ comparisons,
where $g$ is the number of remaining pairs of rows. After one
differential saving is identified, the child row needs to be updated,
which requires at most $2n'$ comparisons. Correspondingly, computing the differential savings relative to the new child row needs up to $2n'(n-2)$ comparisons since the parent row is ineligible. So it will take at most
$\sum_{g=n(n-1)/2}^{1}\bigl(2n'(n-2)+2n'+g-1\bigr)\approx O(n^4+n^2n')$ comparisons. Updating $\boldsymbol{K}$ does not requires extra computation.
Therefore the number of total comparisons for differential
transformations is $O(n^4+n^2n')$.

To initialize $\boldsymbol{R}$, we scan the matrix row by row to find the recurrences of each two-summand pattern. For any row, we increase $R_{ij}$ by one if the two-summand pattern  $X_i+ X_{i + j - 1}$ is present. Since there are at most $n'$ 1's in a row, it has at most $n'(n'-1)/2$ two-summand patterns and hence requires at most $n'(n' - 1) / 2$ additions. Thus Step~\ref{alg:reduction}.4 needs at most $nn'(n'-1)/2$ additions.
For the first recurrence transformation, it will
need at most $(n+n')(n+n'-1)nn'/2$ comparisons to find the greatest in $\boldsymbol{R}$, because there are $(n+n')(n+n'-1)/2$ possible two-summand patterns when all $n$ sums in $\boldsymbol{Y}$ have become summands after differential transformations.
After that, to identify each two-summand
pattern, it needs $(s+n+n')(s+n+n'-1)/2-1$ comparisons, where $s$ is
the number of identified patterns. After a pattern is identified,
all rows with the pattern need to be updated. For each pattern, it
needs to go through at most $n$ rows. Hence it requires at most
$2nn'$ comparisons. If the pattern is forced, it needs to go through
all identified patterns, which requires at most $n's$ comparisons
for one row and $nn's$ comparison for $n$ rows.
It requires at most $nn'(n'-1)/2$ additions to update $\boldsymbol{R}$.
Under direct computation, $\boldsymbol{M}$ requires at most $nn'$ additions.
By identifying one pattern, the number of additions increases by one while saving at least one addition than direct computation.
Based on this observation, we deduce that there are at most $nn'/2$ identified patterns.
Thus the number of comparisons required in Step~\ref{alg:reduction}.5 is at most $\sum_{s=0}^{nn'/2-1}\bigl((s+n+n')(s+n+n'-1)/2-1+2nn'+nn's\bigr)\approx
O(n^3n'^3)$.
The number of required additions is at most $\sum_{s=0}^{nn'/2-1}\bigl(nn'(n'-1)/2\bigr) \approx O(n^2n'^3)$.
Assuming additions have the same complexity as comparisons, it is negligible.
To identify one reversal pattern needs $nn'(n'-1)$,
and its complexity is also negligible compared to those of other parts.
Hence the complexity of our CSE algorithm is $O(n^3n'^3+n^4)$, or
$O(n^6)$ assuming $n=n'$.

The time complexity above is for one run of
Algorithm~\ref{alg:reduction}. Since Algorithm~\ref{alg:reduction}
is probabilistic, it is necessary to run it multiple times to obtain
good results. However, a very large number of runs is not necessary
even for large problems, since the variance between different runs
is relatively small in comparison with the total number of required
additions.

The storage complexity of Algorithm~\ref{alg:reduction} includes
five parts: $\boldsymbol{M}$, $\boldsymbol{D}$, $\boldsymbol{R}$,
$\boldsymbol{K}$, and the list of identified two-summand patterns.
For $\boldsymbol{M}$, it is at most $nn'$. For $\boldsymbol{D}$, it
is $n^2$ and can be reduced to $n(n-1)$ since $\boldsymbol{D}_{ii}$
is not necessary. Since there are at most $nn'/2$ identified
patterns, the storage of $\boldsymbol{R}$ is at most
$(nn'/2+n+n')(nn'/2+n+n'-1)/2$ and it takes at most $nn'$ to keep
the list of identified patterns. The three-dimensional array
$\boldsymbol{K}$ requires at most $n$ times of $\boldsymbol{M}$.
Hence the total storage complexity is at most $O(n^2n'^2)$, or
$O(n^4)$ assuming $n=n'$.

Note that the upper bound $nn'/2$ of the number of identified patterns for an $n \times n'$ matrix is usually not tight.
For example, for a $1023\times1023$ matrix, only less than 30,000 patterns are identified in our simulation.

\section{Relations among Various CFFTs}\label{sec:var}
Our CSE algorithm can be used to reduce the additive complexities of
various CFFTs. In this section, we will investigate their properties
and establish the relations among them. This study also simplifies
the analysis of their multiplicative and additive complexities as
well as performance comparison in Section~\ref{sec:fft}.

Let us first study the properties of a block diagonal matrix
$\boldsymbol{L}=\mathrm{diag}(\boldsymbol{L}_0, \boldsymbol{L}_1,
\dotsc, \boldsymbol{L}_{l-1})$, where $\boldsymbol{L}_i$'s are all
circulant matrices. Clearly, $\boldsymbol{L}_i$'s are all symmetric
and hence $\boldsymbol{L}$ is also symmetric.
We formally present a result mentioned in \cite{Costa04} and
\cite[pp. 273]{Hong93}, which can be proved easily by inspection.
\begin{lemma}
	Given $\boldsymbol{L}=\mathrm{diag}(\boldsymbol{L}_0, \boldsymbol{L}_1, \dotsc, \boldsymbol{L}_{l-1})$ that is a block diagonal matrix where $\boldsymbol{L}_i$'s are all circulant, its inverse $\boldsymbol{L}^{-1}=\mathrm{diag}(\boldsymbol{L}_0^{-1}, \boldsymbol{L}_1^{-1}, \dotsc, \boldsymbol{L}_{l-1}^{-1})$ is also a block diagonal matrix where $\boldsymbol{L}_i^{-1}$'s are all circulant.
Furthermore, suppose $\boldsymbol{L}_i$ is generated by $\gamma_i$ and $\boldsymbol{b}_i=(\gamma_i, \gamma_i^2, \dotsc, \gamma_i^{2^{m_i-1}})$ is a normal basis, then $\boldsymbol{L}_i^{-1}$ is a circulant matrix generated by $\beta _i$, where $(\beta_i, \beta_i^2, \dotsc, \beta_i^{2^{m_i-1}})$ is the dual basis of $\boldsymbol{b}_i$.
\label{lem:cui}
\end{lemma}
Thus, for DCFFTs and SCFFTs $\boldsymbol{L}_i \boldsymbol{f}_i$ is a
cyclic convolution and can be calculated by the bilinear form
$\boldsymbol{Q}_i(\boldsymbol{R}_i\boldsymbol{b}_i
\cdot\boldsymbol{P}_i\boldsymbol{f}_i)=\boldsymbol{Q}_i
(\boldsymbol{c}_i\cdot\boldsymbol{P}_i\boldsymbol{f}_i)$
\cite{Winograd77,Wagh83,Blahut83,Blahut84}, where
$\boldsymbol{b}_i=(\gamma_i, \gamma_i^2, \dotsc,
\gamma_i^{2^{m_i-1}})$. For ICFFTs, by Lemma~\ref{lem:cui}
$\boldsymbol{L}_i^{-1} \boldsymbol{f}_i$ is also a cyclic
convolution given by the bilinear form
$\boldsymbol{Q}_i\bigl(\boldsymbol{R}_i (\beta_i, \beta_i^2, \dotsc,
\beta_i^{2^{m_i-1}})
\cdot\boldsymbol{P}_i\boldsymbol{f}_i\bigr)=\boldsymbol{Q}_i
(\boldsymbol{c}_i^*\cdot\boldsymbol{P}_i\boldsymbol{f}_i)$. There
are different bilinear forms of cyclic convolution and all of them
can be used in CFFTs. Henceforth, we assume that the \textbf{same}
bilinear forms ($\boldsymbol{P}_i$'s and $\boldsymbol{Q}_i$'s) are
used in all CFFTs. In this paper, we focus on the CFFTs with the
following forms:
\begin{align}
		\text{DCFFT} \quad \boldsymbol{F}
					&=\boldsymbol{ALf'}\nonumber\\
					&=\boldsymbol{AQ}(\boldsymbol{c}\cdot \boldsymbol{Pf}') \label{eqn:dir}\\
		\text{SCFFT} \quad \boldsymbol{F}'
					&=\boldsymbol{L}^T\boldsymbol{A}'^{T}\boldsymbol{f}'\nonumber\\
			  	 	 &=\boldsymbol{P}^T\bigl(\boldsymbol{c}\cdot(\boldsymbol{A}'\boldsymbol{Q})^T\boldsymbol{f}'\bigr)\label{eqn:tra}\\
		\text{ICFFT} \quad \boldsymbol{F}''
					&=\boldsymbol{L}^{-1}\boldsymbol{A}^{-1}\boldsymbol{f}\nonumber\\
					&= \boldsymbol{P}^T(\boldsymbol{c^*}\cdot \boldsymbol{Q}^T\boldsymbol{A}^{-1}\boldsymbol{f})\label{eqn:inv}
	\end{align}
where $\boldsymbol{Q}$ and $\boldsymbol{P}$ are binary matrices and usually sparse, and $\boldsymbol{A}$ is a dense binary square matrix.
Note that the equality \eqref{eqn:tra} is due to $\boldsymbol{L} =\boldsymbol{Q} \boldsymbol{C}\boldsymbol{P}$ where $\boldsymbol{C} = \mathrm{diag}(c_0, c_1, \dotsc, c_{n-1})$;
the equality \eqref{eqn:inv} follows \eqref{eqn:tra} and is a direct application of Lemma~\ref{lem:cui}.
Due to the symmetric properties of $\boldsymbol{L}$ and $\boldsymbol{L}^{-1}$, the above CFFTs have alternative forms: DCFFTs are also given by $\boldsymbol{F}=\boldsymbol{A}\boldsymbol{P}^T(\boldsymbol{c}\cdot \boldsymbol{Q^Tf}')$;
SCFFTs are also given by $\boldsymbol{F}' =\boldsymbol{Q}\bigl(\boldsymbol{c}\cdot(\boldsymbol{A}'\boldsymbol{Q}^T)^T\boldsymbol{f}'\bigr)$;
ICFFTs are also given by $\boldsymbol{F}''= \boldsymbol{Q}(\boldsymbol{c^*}\cdot \boldsymbol{P}\boldsymbol{A}^{-1}\boldsymbol{f})$.
However, these alternative forms can be considered as the forms in \eqref{eqn:dir}, \eqref{eqn:tra}, and \eqref{eqn:inv} with different $\boldsymbol{P}$ and $\boldsymbol{Q}$ matrices.
Since we assume all the bilinear forms are the same, we will not consider the alternative forms further.

We observe that all CFFTs in \eqref{eqn:dir}, \eqref{eqn:tra}, and
\eqref{eqn:inv} are determined by two factors. First, they all
depend on the order of cyclotomic cosets, i.e., the coset leaders
$k_i$'s, which in turn determine the coset size $m_i$'s. As in
\cite{Trifonov03}, we assume the same normal basis is used for all
cyclotomic cosets of the same size. Hence, all CFFTs also depend on
the normal basis selected for each subfield $\GF(2^{m_i})$. For
simplicity, we denote the collections of DCFFTs, SCFFTs, and ICFFTs
for different $k_i$'s, $m_i$'s and the normal bases as
$\mathcal{D}$, $\mathcal{S}$, and $\mathcal{I}$, respectively.
Next, we investigate the impact on computational complexities of
CFFTs by the two factors above. We will consider first
multiplicative complexities and then additive complexities.

\begin{lemma}
Assuming that the same bilinear forms are used, DCFFTs, SCFFTs, and ICFFTs as defined in \eqref{eqn:dir}, \eqref{eqn:tra}, and \eqref{eqn:inv} have the same multiplicative complexities.
\label{lem:mul}
\end{lemma}
\begin{IEEEproof}
The multiplicative complexity is determined by the number of non-one
entries in $\boldsymbol{c}$ in DCFFTs and SCFFTs or $\boldsymbol{c}^*$
in ICFFTs (all elements in $\boldsymbol{c}$ or $\boldsymbol{c}^*$ are
non-zero). Since using normal bases, the number of 1's in
$\boldsymbol{c}$ and $\boldsymbol{c}^*$ are both the number of
all-one rows in all $\boldsymbol{R}_i$'s. Thus the multiplicative complexity is
independent of the choices of normal bases and independent of the
constant vectors $\boldsymbol{c}$ or $\boldsymbol{c}^*$.
\end{IEEEproof}

The additive complexities of all CFFTs are due to the matrix-vector
multiplications needed in CFFTs. Clearly, the number of additions
required to compute any matrix-vector multiplication $\boldsymbol{Y}=\boldsymbol{M}\boldsymbol{X}$
varies with the implementation. In the following, we will consider
additive complexities under direct computation. As pointed out in Section~\ref{sec:ralg}, to compute
$\boldsymbol{Y}=\boldsymbol{MX}$ by direct computation, it needs
$W(\boldsymbol{M}) - n$ additions.
In some cases the additive complexities of two
matrix-vector multiplications can be related regardless of
implementation. We say two matrix-vector multiplications are
\emph{additively equivalent} if one matrix-vector multiplication can achieve any
additive complexity the other can, and vice versa. An important case
of additive equivalence is given in the following lemma without proof.
\begin{lemma}\label{lem:per}
If two binary matrices $\boldsymbol{M}$ and $\boldsymbol{M}'$
satisfy
$\boldsymbol{M}'=\boldsymbol{\Pi}\boldsymbol{M}\boldsymbol{\Pi}'$,
where $\boldsymbol{\Pi}$ and $\boldsymbol{\Pi}'$ are two permutation
matrices, then the matrix-vector multiplications defined by
$\boldsymbol{M}$ and $\boldsymbol{M}'$ are additively equivalent.
\end{lemma}

With a slight abuse of terminology, we
say two CFFTs are \emph{additively equivalent} when their
corresponding matrices are additively equivalent.
By a straightforward proof, we have the following property:
\begin{lemma}
For any two CFFTs in $\mathcal{D}$ that differ \textbf{only} in
$k_i$'s and $m_i$'s, their $\boldsymbol{A}$'s and $\boldsymbol{L}$'s are additively
equivalent, respectively. Thus, the two CFFTs in $\mathcal{D}$ are
additively equivalent. The same property holds for $\mathcal{S}$ and
$\mathcal{I}$.\label{lem:cosets}
\end{lemma}

We now consider additive complexities for all CFFTs when normal
bases vary, too.
\begin{lemma}
All CFFTs in $\mathcal{D}$ have the same additive complexity under
direct computation. So do those in $\mathcal{S}$ and $\mathcal{I}$,
respectively.\label{lem:same}
\end{lemma}
\begin{IEEEproof}
It suffices to prove the first part, and the arguments for
$\mathcal{S}$ and $\mathcal{I}$ are similar. First, since different
orders of cosets result in additively equivalent DCFFTs due to
Lemma~\ref{lem:cosets}, we assume the same order of cosets and
consider only different normal bases without loss of generality.
Realizing that different normal bases would not change
$\boldsymbol{P}$ and $\boldsymbol{Q}$ in \eqref{eqn:dir}, we focus
on how different normal bases impact $\boldsymbol{A}\boldsymbol{Q}$.
Expressing $\boldsymbol{A}$ as
$[\boldsymbol{A}_0\mid\boldsymbol{A}_1\mid\dotsb\mid\boldsymbol{A}_{l-1}]$
where $\boldsymbol{A}_i$ is a $(2^m-1)\times m_i$ binary matrix,
$\boldsymbol{F}=\boldsymbol{A}\boldsymbol{L}\boldsymbol{f}'=[\boldsymbol{A}_0\boldsymbol{L}_0\mid\boldsymbol{A}_1\boldsymbol{L}_1\mid\dotsb\mid\boldsymbol{A}_{l-1}\boldsymbol{L}_{l-1}]\boldsymbol{f}'$.
For each $\boldsymbol{A}_i$, the rows are $(2^m-1)/(2^{m_i}-1)$
copies of the set of $m_i$-bit row vectors with all combinations except all zeros. Thus
$\boldsymbol{A}_i$'s corresponding to different normal bases in
$\GF(2^{m_i})$ are equivalent up to permutation. Recall that
$\boldsymbol{Q}$ is a block matrices for which the blocks off the
diagonal are zero matrices and the diagonal blocks are
$\boldsymbol{Q}_i$'s.
Thus,
$\boldsymbol{A}\boldsymbol{Q}=[\boldsymbol{A}_0\boldsymbol{Q}_0\mid\boldsymbol{A}_1\boldsymbol{Q}_1\mid\dotsb\mid\boldsymbol{A}_{l-1}\boldsymbol{Q}_{l-1}]$.
Thus $\boldsymbol{A}\boldsymbol{Q}$'s corresponding to different normal bases also have the same additive complexity under direct computation.
Hence all DCFFTs in $\mathcal{D}$ have the same additive complexity under direct computation.
\end{IEEEproof}

From Lemma~\ref{lem:cui}, we establish a relation between
$\mathcal{I}$ and $\mathcal{S}$.
	\begin{lemma}
Given an ICFFT $\boldsymbol{F}''=\boldsymbol{L}^{-1}\boldsymbol{A}^{-1}\boldsymbol{f}$, there exists an SCFFT
$\boldsymbol{F}'=\boldsymbol{L}'\boldsymbol{A}'^T\boldsymbol{f}'$ such that $\boldsymbol{L}'=\boldsymbol{L}^{-1}$ and $\boldsymbol{A}'^T$ and
$\boldsymbol{A}^{-1}$ are equivalent up to permutation, and vice
versa.\label{lem:eqi}
	\end{lemma}
\begin{IEEEproof}
It suffices to show the first part, and the argument for the second part is similar.
For a DCFFT given by $\boldsymbol{F} =
\boldsymbol{A}\boldsymbol{L}\boldsymbol{\Pi}\boldsymbol{f}$, the
transform $\boldsymbol{F}^* =
\boldsymbol{\Pi}^{-1}\boldsymbol{L}^{-1}\boldsymbol{A}^{-1}\boldsymbol{f}$
is another DFT, where $\boldsymbol{F}^*= (F_0, F_{n-1}, F_{n-2},
\dotsc, F_1)=\boldsymbol{\Pi}^*\boldsymbol{F}$ and
$\boldsymbol{\Pi}^*$ is a permutation matrix. Given an ICFFT
$\boldsymbol{F}''=\boldsymbol{L}^{-1}\boldsymbol{A}^{-1}\boldsymbol{f}$,
clearly
$\boldsymbol{F}''=\boldsymbol{\Pi}\boldsymbol{F}^*=\boldsymbol{\Pi}\boldsymbol{\Pi}^*\boldsymbol{F}$.
Suppose the indices of the components of $\boldsymbol{F}'=\boldsymbol{\Pi}\boldsymbol{F}$ are in the order as $(k_0, k_0 2, \dotsc, k_0 2^{m_0 - 1}, \dotsc,
k_{l - 1} 2^{m_{l - 1} - 1}) \bmod n$, then the indices of the
components of $\boldsymbol{F}''=\boldsymbol{\Pi}\boldsymbol{F}^*$
are in the order as $(n - k_0, n - k_0 2, \dotsc, n -
k_0 2^{m_0 - 1}, \dotsc,
n - k_{l - 1} 2^{m_{l - 1} - 1})\bmod n$. Note that both modulo operations above
are componentwise. Since $n-k_i2^j \equiv (n-k_i)2^j \bmod n$,
$\boldsymbol{F}''$ is also ordered in cyclotomic cosets. Let us
consider an SCFFT with the same order of cyclotomic cosets:
$\boldsymbol{F}''=\boldsymbol{L}''\boldsymbol{A}''^T\boldsymbol{f}''$
where
$\boldsymbol{f}''=\boldsymbol{\Pi}\boldsymbol{\Pi}^*\boldsymbol{f}$.
Note that the order of the cyclotomic cosets sizes $m_i$ remains the
same in $\boldsymbol{L}''$ and $\boldsymbol{L}^{-1}$. Thus by
Lemma~\ref{lem:cui} there exist normal bases such that
$\boldsymbol{L}''=\boldsymbol{L}^{-1}$. Choosing such normal bases,
we construct an SCFFT $\boldsymbol{F}'' =
\boldsymbol{L}^{-1}\boldsymbol{A}''^T\boldsymbol{f}''=\boldsymbol{L}^{-1}\boldsymbol{A}^{-1}\boldsymbol{f}$.
Thus
$\boldsymbol{L}^{-1}(\boldsymbol{A}''^T\boldsymbol{\Pi}\boldsymbol{\Pi}^*-\boldsymbol{A}^{-1})\boldsymbol{f}=0$
for arbitrary $\boldsymbol{f}$ and full rank $\boldsymbol{L}^{-1}$.
Hence
$\boldsymbol{A}^{-1}=\boldsymbol{A}''^T\boldsymbol{\Pi}^*\boldsymbol{\Pi}$.
\end{IEEEproof}
Note that Lemma~\ref{lem:eqi} holds \textbf{regardless of implementation}.
Since this mapping exists for any ICFFTs or SCFFTs,
Lemma~\ref{lem:eqi} implies that ICFFTs and SCFFTs are additively
equivalent.

Finally, we are ready to relate the additive complexities of all CFFTs under direct computation.
\begin{lemma}
The DCFFTs, SCFFTs, and ICFFTs in \eqref{eqn:dir}, \eqref{eqn:tra}, and \eqref{eqn:inv} all have the same additive complexity under direct computation.
\label{lem:add}
\end{lemma}
\begin{IEEEproof}
Due to Lemma~\ref{lem:same}, it is sufficient to show that the
additive complexities of two CFFTs of different types are the same,
which holds for an SCFFT and an ICFFT by Lemmas~\ref{lem:same} and
\ref{lem:eqi}. Now let us show it is the same for a DCFFT and an SCFFT\@.

In length-$n$ DCFFTs, $\boldsymbol{A}$ is an $n \times n$ matrix,
$\boldsymbol{Q}$ is an $n \times n'$ matrix ($n'>n$), and
$\boldsymbol{P}$ is an $n' \times n$ matrix. Under direct
computation, the number of required additions for a DCFFT defined in
\eqref{eqn:dir} is $W(\boldsymbol{AQ})-n+W(\boldsymbol{P})-n'$.
Since $\boldsymbol{f}'=\boldsymbol{\Pi f}$, we have
$\boldsymbol{F}'=\boldsymbol{A}'\boldsymbol{Q}(\boldsymbol{c}\cdot
\boldsymbol{Pf}')$, where $\boldsymbol{F}'=\boldsymbol{\Pi F}$ and
$\boldsymbol{A}'=\boldsymbol{\Pi A}$. For an SCFFT $\boldsymbol{F}'
=
\boldsymbol{P}^T(\boldsymbol{c}\cdot(\boldsymbol{A}'\boldsymbol{Q})^T\boldsymbol{f}')$,
 the additive complexity under direct computation is
$W\bigl((\boldsymbol{A}'\boldsymbol{Q})^T\bigr) - n' + W(\boldsymbol{P}^T)
-n$. Since $\boldsymbol{A}'\boldsymbol{Q} =
\boldsymbol{\Pi}\boldsymbol{AQ}$, so $\boldsymbol{A}'\boldsymbol{Q}$
and $\boldsymbol{AQ}$ have the same number of 1's. Since matrix
transpose does not change the number of 1's,
$W\bigl((\boldsymbol{A}'\boldsymbol{Q})^T\bigr) = W(\boldsymbol{AQ})$ and
$W(\boldsymbol{P}^T)=W(\boldsymbol{P})$. Hence any DCFFTs in
\eqref{eqn:dir} and any SCFFTs in \eqref{eqn:tra} have the same
additive complexity under direct computation. An alternative direct
computation for both DCFFTs and SCFFTs is to multiply
$\boldsymbol{A}$ and $\boldsymbol{Q}$ separately. It is easy to
verify that the conclusion is the same.
\end{IEEEproof}

\section{CFFTs with Reduced Additive Complexities}\label{sec:fft}
Using Algorithm~\ref{alg:reduction}, we construct CFFTs with reduced
additive complexities for lengths $2^m-1$ up to 1023, and we present
their complexities in Table~\ref{tab:fcfft}. CFFTs of length beyond
1023 are not considered because for two reasons: first,
lengths beyond 1023 are rarely needed
for the primary application considered in this paper, Reed--Solomon decoders;
second, efficient cyclic convolutions for CFFTs of longer lengths (for example, 11-point cyclic convolution for length-2047 CFFTs) are not available in \cite{Blahut83,Blahut84,Trifonov}.
For all our CFFTs, the cyclotomic cosets are ordered by their leaders;
for cyclic convolutions of lengths up to nine, we use the bilinear forms provided in \cite{Trifonov}, and we construct a length-10 cyclic convolution based on those of lengths two and five, by the Agarwal--Cooley algorithm \cite{Agarwal77};
the primitive polynomials and vector-space representations in \cite[Sec. B.3]{Wicker95} are used for all fields;
for each field, we choose the normal basis whose leader is the smallest power of the primitive element.
We observe that the multiplicative complexities are the same for all CFFTs due to Lemma~\ref{lem:mul}.
Due to Lemma~\ref{lem:eqi}, SCFFTs and ICFFTs are additively equivalent, and the additive complexities of both SCFFTs and ICFFTs are presented together in Table~\ref{tab:fcfft}.
We also observe that SCFFTs and ICFFTs require more additions than DCFFTs, and the reason for this was given in \cite{Fedorenko06}.

	\begin{table}[htpb]
		\centering
		\caption{Complexities of Full Cyclotomic FFTs}
		\label{tab:fcfft}
		\begin{tabular}{|c|c|c|c|c|c|c|}
				\hline
				\multirow{3}{*}{$n$} & \multirow{3}{*}{Mult.} & \multicolumn{5}{c|}{Additions}\\
				\cline{3-7}
		& & \multicolumn{2}{c|}{DCFFT} & \multicolumn{3}{c|}{SCFFT/ICFFT}\\
				\cline{3-7}
		& & Ours & \cite{Trifonov03} & Ours & \cite{Costa04} & \cite{Fedorenko06}\\
				\hline
				7 & 6 & 24 & 25 & 24 & 24 & -\\
				\hline
				15 & 16 & 74 & 77 & 76 & - & 91\\
				\hline
				31 & 54 & 299 & 315 & 307 & - & -\\
				\hline
				63 & 97 & 759 & 805 & 804 & - & -\\
				\hline
				127 & 216 & 2576 & 2780 & 3117 & - & -\\
				\hline
				255 & 586 & 6736 & 7919 & 6984 & - & -\\
				\hline
				511 & 1014 & 23130 & 26643 & 27192 & - &\\
				\hline
				1023 & 2827 & 75360 & - & 77276 & - & -\\
				\hline
		\end{tabular}
	\end{table}

\newcommand{\customizedtable}{
\begin{table*}[htbp!]
		\centering
		\caption{Complexities of Full FFTs}
		\label{tab:ffft}
	\small\addtolength{\tabcolsep}{-3pt}
	\scalebox{
	\ifCLASSOPTIONtwocolumn
	0.87
	\else
	0.73
	\fi
	}{
		\begin{tabular}{|c|c|c|c|c|c|c|c|c||c|c|c||c|c|c||c|c|c||c|}
				\hline
				\multirow{2}{*}{$n$} & \multicolumn{2}{c|}{Horner's rule} & \multicolumn{2}{c|}{Goertzel's alg.} & \multicolumn{2}{c|}{\cite{Zakharova92}} & \multicolumn{3}{c|}{\cite{Wang88}} & \multicolumn{3}{c|}{Bergland's alg.} & \multicolumn{3}{c|}{Prime-factor \cite{Truong06}} & \multicolumn{3}{c|}{Our DCFFTs}\\
				\cline{2-19}
				& Mult. & Add. & Mult. & Add. & Mult. & Add. & Mult. & Add. & Total & Mult. & Add. & Total & Mult. & Add. & Total & Mult. & Add. & Total\\
				\hline
				7 & 36 & 42 & 12 & 42 & 6 & 26 & 29 & 29 & 174 & - & - & - & 9 & 37 & 82 & 6 & 24 & 54\\
				\hline
				15 & 196 & 210 & 38 & 210 & 16 & 100 & 41 & 97 & 384 & - & - & - & - & - & - & 16 & 74 & 186\\
				\hline
				31 & 900 & 930 & 120 & 930 & 60 & 388 & 289 & 289 & 2890 & - & - & - & 108 & 612 & 1584 & 54 & 299 & 785\\
				\hline
				63 & 3844 & 3906 & 282 & 3906 & 97 & 952 & 801 & 801 & 9612 & - & - & - & - & - & - & 97 & 759 & 1826\\
				\hline
				127 & 15876 & 16002 & 756 & 16002 & 468 & 3737 & 2113 & 2113 & 29582 & - & - & - & - & - & - & 216 & 2576 & 5384\\
				\hline
				255 & 64516 & 64770 & 1718 & 64770 & 646 & 35503 & 1665 & 5377 & 30352 & 5610 & 5610 & 89760 & 1135 & 3887 & 20902 & 586 & 6736 & 15526\\
				\hline
				511 & 260100 & 260610 & 4044 & 260610 & - & - & 13313 & 13313 & 239634 & 39858 & 39858 & 717444 & 6516 & 17506 & 128278 & 1014 & 23130 & 40368\\
				\hline
				1023 & 1044484 & 1045506 & 9032 & 1045506 & - & - & 32257 & 32257 & 645140 & 42966 & 42966 & 859320 & 5915 & 30547 & 142932 & 2827 & 75360 & 129073\\
				\hline
		\end{tabular}
		}
	\end{table*}
	}
\ifCLASSOPTIONtwocolumn
\customizedtable
	\fi

In Table~\ref{tab:fcfft}, we also compare the additive complexities
of our CFFTs to
those in \cite{Trifonov03,Costa04,Fedorenko06}, the best
results of CFFTs in the open literature to our knowledge\footnote{A
length-15 DCFFT with 76 additions was reported in~\cite{Trifonov}.}.
In Table~\ref{tab:fcfft}, some entries are blank due to
unavailability of comparable data: the additive complexity of DCFFT
of length 1023 is not provided in \cite{Trifonov03};
only length-7
ICFFT was provided in \cite{Costa04} and only length-15 SCFFTs was
provided in \cite{Fedorenko06}.
For length-7 FFT, both our DCFFT and SCFFT achieve the smallest additive complexity of the ICFFT in \cite{Costa04};
for lengths 15, 31, 63, and 127, our CFFTs have additive complexities 4\%, 5\%, 6\%, and 7\% smaller than those reported in \cite{Trifonov03};
for lengths 255 and 511, our CFFTs reduce additive complexities by 15\% and 13\%, respectively, than their counterparts in \cite{Trifonov03}.
To compare our length-7 DCFFT with that in \cite{Trifonov03}, see Appendix~\ref{sec:dc7}.

We also compare our results to other FFT algorithms in
Table~\ref{tab:ffft}. For Horner's rule \cite{Wicker94}, Goertzel's
algorithm \cite{Blahut83}, Zakharova's method \cite{Zakharova92},
the complexities are reproduced from \cite{Trifonov03} except that
the complexities of length-1023 FFTs are reproduced from
\cite{Truong06};
the complexities of Bergland's algorithm
\cite{Brigham88} and the prime-factor FFTs \cite{Truong06} are
obtained from \cite{Truong06,Truong06a}. For reference, we also
consider the algorithm proposed by Wang and Zhu \cite{Wang88}, which
is known to be asymptotically fast, and its complexities are
obtained from \cite[eq. (11) and (12)]{Wang88}.
\ifCLASSOPTIONonecolumn
\customizedtable
	\fi

Since all the algorithms require both multiplicative and additive
complexities, it is clear that a metric for the total complexities
is needed for comparison. We use a weighted sum of the additive and
multiplicative complexities as the metric, assuming the complexity
of each multiplication is $2m-1$ times as that of an addition. Our
assumption is based on both hardware and software considerations. In
hardware implementation, a multiplier over $\mathrm{GF}(2^m)$
generated by trinomials requires $m^2-1$ XOR and $m^2$ AND gates
(see, e.g., \cite{Sunar99}), while an adder requires $m$ XOR gates.
Assuming that XOR and AND gates require the same area, the area
complexity of a field multiplier is $2m$ times that of an adder over
$\mathrm{GF}(2^m)$.
In software implementation, the complexity can
be measured by the number of word-level operations (see, for
example, \cite{Mahboob05}). Using the shift and add method as in
\cite{Mahboob05}, a multiplication requires $m-1$ shift and $m$ XOR
word-level operations, respectively while an addition needs only one
XOR word-level operation. Whenever the complexity of a
multiplication is more than $2m-1$ times as complex as that of an
addition (for example, in the hardware implementation described
above), our assumption above underestimates the relative complexity
of multiplications and hence puts our results in a disadvantage in
comparison to other FFT algorithms since CFFTs have reduced
multiplicative complexities. We would also like to point out the
similarity between our metric and the one used in \cite{Wang88},
where the multiplication over $\GF(2^m)$ was treated $2m$ times
as complex as an addition.

The total complexities of Horner's rule, Goertzel's algorithm, and
\cite{Zakharova92} are not presented in Table~\ref{tab:ffft} since
the advantage in complexities of our CFFTs over Horner's rule,
Goertzel's algorithm, and \cite{Zakharova92} is clear: our CFFTs
require fewer multiplications \textbf{and} fewer additions;
the
savings achieved by our CFFTs are very significant, and in some
cases the multiplicative complexities of our CFFTs are only small
fractions of other algorithms. We remark that the multiplicative
complexities of Zakharova's method are closer to those of CFFTs,
which is not surprising given their similarities \cite{Trifonov03}.
The total complexities of \cite{Wang88}, Bergland's algorithm, the prime-factor FFTs \cite{Truong06} and our CFFTs are presented in Table~\ref{tab:ffft}, since in comparison to these algorithms our CFFTs have smaller multiplicative complexities but higher additive complexities.
In comparison to \cite{Wang88}, our CFFTs achieve total complexity savings of 69\%, 52\%, 73\%, 81\%, 82\%, 49\%, 83\% and 80\% for lengths $7, 15, \dotsc, 1023$, respectively.
For lengths 255, 511, and 1023, our CFFTs achieve total complexity savings of 83\%, 94\%, and 85\% over Bergland's algorithm, and 26\%, 69\%, and 10\% over the prime-factor FFTs \cite{Truong06}, respectively.

We remark that, as in many previous works (see, for example, \cite{Trifonov03,Costa04,Fedorenko06}), only the multiplications and additions are considered in the complexity comparison. This is reasonable if the CFFTs are implemented by combinational logic, and the required numbers of multiplications and additions translate to the numbers of finite field multipliers and adders in combinational logic. Under the same assumption, memory overhead and intermediate memory access are not considered in the comparison above.  This would not be the case if CFFTs were implemented in software, but this is beyond the scope of this paper.

\section*{Acknowledgment}
The authors would like to thank Prof. P. Trifonov for providing details of CFFTs.
They are grateful to Prof. M.D. Wagh for introducing them to his fast convolution algorithms.
They would also thank Prof. P.D. Chen for valuable discussions.
The authors would also like to thank the reviewers for their constructive comments, which have resulted in significant improvements in the manuscript.
\appendices

\section{Length-7 DCFFT}\label{sec:dc7}

	\begin{itemize}
		\item Pre-additions $\boldsymbol{p} = (p_0, p_1, \dotsc, p_8)^T = \boldsymbol{Pf'}$ require 8 additions: $p_0 = f_0$, $p_2 = f_2 + f_4$, $p_3 = f_1 + f_2$, $p_4 = f_1 + f_4$,  $p_1= p_2 + f_1$, $p_6= f_6 + f_5$,
$p_7 = f_3 + f_6$, $p_8 = f_3 + f_5$, and $p_5 = p_6 + f_3$.
			\item Pointwise multiplications $\boldsymbol{g} = (g_0, g_1, \dotsc, g_8)^T = \boldsymbol{c} \cdot \boldsymbol{p}$, where
$\boldsymbol{c} = (1, 1, \alpha, \alpha^2, \alpha^4, 1, \alpha, \alpha^2, \alpha^4)^T$,  need 6 multiplications
		\item Post-additions $\boldsymbol{F} = (F_0, F_1, \dotsc, F_6)^T =\boldsymbol{AQg}$ require 16 additions: $t_0 = g_3 + g_4$, $t_1 = g_0 + g_1$, $t_2 = g_1 + g_5$, $F_0 = g_0 + t_2$,
$t_3 = g_2 + g_4$, $t_4 = g_8 + t_3$, $t_5 = g_7 + t_4$, $F_5 = t_1 + t_5$,
$t_6 = g_6 + t_4$, $t_7 = t_1 + t_6$, $F_6 = t_0 + t_7$, $F_3 = F_6 + t_5$,
$t_8 = t_3 + t_2$, $F_2 = F_3 + t_8$, $F_1 = F_2 + t_6$, and $F_4 = t_2 + t_7$.
	\end{itemize}

\bibliographystyle{IEEEtran}
\bibliography{IEEEabrv,rs}

\begin{thebibliography}{10}
\providecommand{\url}[1]{#1}
\csname url@samestyle\endcsname
\providecommand{\newblock}{\relax}
\providecommand{\bibinfo}[2]{#2}
\providecommand{\BIBentrySTDinterwordspacing}{\spaceskip=0pt\relax}
\providecommand{\BIBentryALTinterwordstretchfactor}{4}
\providecommand{\BIBentryALTinterwordspacing}{\spaceskip=\fontdimen2\font plus
\BIBentryALTinterwordstretchfactor\fontdimen3\font minus
  \fontdimen4\font\relax}
\providecommand{\BIBforeignlanguage}[2]{{%
\expandafter\ifx\csname l@#1\endcsname\relax
\typeout{** WARNING: IEEEtran.bst: No hyphenation pattern has been}%
\typeout{** loaded for the language `#1'. Using the pattern for}%
\typeout{** the default language instead.}%
\else
\language=\csname l@#1\endcsname
\fi
#2}}
\providecommand{\BIBdecl}{\relax}
\BIBdecl

\bibitem{Blahut79}
R.~E. Blahut, ``Transform techniques for error control codes,'' \emph{IBM J.
  Res. Develop.}, vol.~23, no.~3, pp. 299--315, May 1979.

\bibitem{Truong06}
T.~K. Truong, P.~D. Chen, L.~J. Wang, I.~S. Reed, and Y.~Chang, ``Fast, prime
  factor, discrete {F}ourier transform algorithms over $\mathrm{GF}(2^m)$ for
  $8 \le m \le 10$,'' \emph{Inf. Sci.}, vol. 176, no.~1, pp. 1--26, Jan. 2006.

\bibitem{Truong06a}
T.~K. Truong, P.~D. Chen, L.~J. Wang, and T.~C. Cheng, ``Fast transform for
  decoding both errors and erasures of {R}eed--{S}olomon codes over
  $\mathrm{GF}(2^m)$ for $8 \le m \le 10$,'' \emph{{IEEE} Trans. Commun.},
  vol.~54, no.~2, pp. 181--186, Feb. 2006.

\bibitem{Lin07}
T.-C. Lin, T.~K. Truong, and P.~D. Chen, ``A fast algorithm for the syndrome
  calculation in algebraic decoding of {R}eed--{S}olomon codes,'' \emph{{IEEE}
  Trans. Commun.}, vol.~55, no.~12, pp. 1--5, Dec. 2007.

\bibitem{Zakharova92}
T.~G. Zakharova, ``Fourier transform evaluation in fields of characteristic
  2,'' \emph{Probl. Inf. Transm.}, vol.~28, no.~2, pp. 154--167, 1992.

\bibitem{Trifonov03}
\BIBentryALTinterwordspacing
P.~V. Trifonov and S.~V. Fedorenko, ``A method for fast computation fo the
  {F}ourier transform over a finite field,'' \emph{Probl. Inf. Transm.},
  vol.~39, no.~3, pp. 231--238, 2003. [Online]. Available:
  \url{http://dcn.infos.ru/~petert/papers/fftEng.pdf}
\BIBentrySTDinterwordspacing

\bibitem{Costa04}
\BIBentryALTinterwordspacing
E.~Costa, S.~V. Fedorenko, and P.~V. Trifonov, ``On computing the syndrome
  polynomial in {R}eed--{S}olomon decoder,'' \emph{Euro. Trans. Telecomms.},
  vol.~15, no.~4, pp. 337--342, 2004. [Online]. Available:
  \url{http://dcn.infos.ru/~petert/papers/syndromes_ett.pdf}
\BIBentrySTDinterwordspacing

\bibitem{Fedorenko06}
S.~V. Fedorenko, ``A method of computation of the discrete {F}ourier transform
  over a finite field,'' \emph{Probl. Inf. Transm.}, vol.~42, no.~2, pp.
  139--151, 2006.

\bibitem{Garey79}
M.~R. Garey and D.~S. Johnson, \emph{Computers and Intractability: A Guide to
  the Theory of NP-Completeness}, ser. Books in the Mathematical
  Sciences.\hskip 1em plus 0.5em minus 0.4em\relax San Francisco, CA: W. H.
  Freeman, 1979.

\bibitem{Cappello84}
P.~Cappello and K.~Steiglitz, ``Some complexity issues in digital signal
  processing,'' \emph{{IEEE} Trans. Acoust., Speech, Signal Process.}, vol.~32,
  no.~5, pp. 1037--1041, Oct. 1984.

\bibitem{Horn85}
R.~A. Horn and C.~R. Johnson, \emph{Matrix Analysis}.\hskip 1em plus 0.5em
  minus 0.4em\relax New York, NY: Cambridge Univ. Press, 1985.

\bibitem{Winograd77}
S.~Winograd, ``Some bilinear forms whose multiplicative complexity depends on
  the field of constants,'' \emph{Math. Syst. Theory}, vol.~10, no.~1, pp.
  169--180, 1977.

\bibitem{Wagh83}
M.~D. Wagh and S.~D. Morgera, ``A new structured design method for convolutions
  over finite fields, {P}art {I},'' \emph{{IEEE} Trans. Inf. Theory}, vol.~29,
  no.~4, pp. 583--595, Jul. 1983.

\bibitem{Blahut83}
R.~E. Blahut, \emph{Theory and Practice of Error Control Codes}.\hskip 1em plus
  0.5em minus 0.4em\relax Reading, MA: Addison-Wesley, 1983.

\bibitem{Blahut84}
------, \emph{Fast Algorithms for Digital Signal Processing}.\hskip 1em plus
  0.5em minus 0.4em\relax Reading, MA: Addison-Wesley, 1984.

\bibitem{Trifonov}
P.~Trifonov, private communication.

\bibitem{Trifonov07}
\BIBentryALTinterwordspacing
------, ``Matrix-vector multiplication via erasure decoding,'' in \emph{Proc.
  XI Int. Symp. Problems of Redundancy in Information and Control Systems},
  Saint-Petersburg, Russia, Jul. 2007, pp. 104--108. [Online]. Available:
  \url{http://dcn.infos.ru/~petert/papers/mo.pdf}
\BIBentrySTDinterwordspacing

\bibitem{Aho74}
A.~V. Aho, J.~E. Hopcroft, and J.~D. Ullman, \emph{The Design and Analysis of
  Computer Algorithms}.\hskip 1em plus 0.5em minus 0.4em\relax Reading, MA:
  Addison-Wesley, 1974.

\bibitem{Potkonjak96}
M.~Potkonjak, M.~B. Srivastava, and A.~P. Chandrakasan, ``Multiple constant
  multiplications: efficient and versatile framework and algorithms for
  exploring common subexpression elimination,'' \emph{{IEEE} Trans.
  Comput.-Aided Design Integr. Circuits Syst.}, vol.~15, no.~2, pp. 151--165,
  Feb. 1996.

\bibitem{Pasko99}
R.~Pa\v{s}ko, P.~Schaumont, V.~Derudder, S.~Vernalde, and
  D.~\v{D}ura\v{c}kov\'a, ``A new algorithm for elimination of common
  subexpressions,'' \emph{{IEEE} Trans. Comput.-Aided Design Integr. Circuits
  Syst.}, vol.~18, no.~1, pp. 58--68, Jan. 1999.

\bibitem{Hartley96}
R.~I. Hartley, ``Subexpression sharing in filters using canonic signed digit
  multipliers,'' \emph{{IEEE} Trans. Circuits Syst. {II}}, vol.~43, no.~10, pp.
  677--688, Oct. 1996.

\bibitem{Martinez-Peiro02}
M.~Mart\'inez-Peir\'o, E.~I. Boemo, and L.~Wanhammar, ``Design of high-speed
  multiplierless filters using a nonrecursive signed common subexpression
  algorithm,'' \emph{{IEEE} Trans. Circuits Syst. {II}}, vol.~49, no.~3, pp.
  196--203, Mar. 2002.

\bibitem{Mahesh08}
R.~Mahesh and A.~P. Vinod, ``A new common subexpression elimination algorithm
  for realizing low-complexity higher order digital filters,'' \emph{{IEEE}
  Trans. Comput.-Aided Design Integr. Circuits Syst.}, vol.~27, no.~2, pp.
  217--229, Feb. 2008.

\bibitem{Dempster95}
A.~G. Dempster and M.~D. Macleod, ``Use of minimum-adder multiplier blocks in
  {FIR} digital filters,'' \emph{{IEEE} Trans. Circuits Syst. {II}}, vol.~42,
  no.~9, pp. 569--577, Sep. 1995.

\bibitem{Chang08}
C.-H. Chang, J.~Chen, and A.~P. Vinod, ``Information theoretic approach to
  complexity reduction of {FIR} filter design,'' \emph{{IEEE} Trans. Circuits
  Syst. {I}}, to be published.

\bibitem{Zhang02}
X.~Zhang and K.~K. Parhi, ``Implementation approaches for the {A}dvanced
  {E}ncryption {S}tandard algorithm,'' \emph{{IEEE} Circuits Syst. Mag.},
  vol.~2, no.~4, pp. 24--46, 2002.

\bibitem{Macleod04}
M.~D. Macleod and A.~G. Dempster, ``Common subexpression elimination algorithm
  for low-cost multiplierless implementation of matrix multipliers,''
  \emph{Electron. Lett.}, vol.~40, no.~11, pp. 651--652, May 2004.

\bibitem{Gustafsson07}
O.~Gustafsson and M.~Olofsson, ``Complexity reduction of constant matrix
  computations over the binary field,'' in \emph{Proc. Int. Workshop Arithmetic
  Finite Fields (WAIFI'07)}, ser. Lecture Notes in Computer Science, vol. 4547,
  Madrid, Spain, Jun. 2007, pp. 103--115.

\bibitem{Bertsekas95}
D.~P. Bertsekas, \emph{Dynamic Programming and Optimal Control}.\hskip 1em plus
  0.5em minus 0.4em\relax Belmont, MA: Athena Scientific, 1995.

\bibitem{Hong93}
J.~Hong and M.~Vetterli, ``Computing $m$ {DFT}'s over $\mathrm{GF}(q)$ with one
  {DFT} over $\mathrm{GF}(q^m)$,'' \emph{{IEEE} Trans. Inf. Theory}, vol.~49,
  no.~1, pp. 271--274, Jan. 1993.

\bibitem{Agarwal77}
R.~Agarwal and J.~Cooley, ``New algorithms for digital convolution,''
  \emph{{IEEE} Trans. Acoust., Speech, Signal Process.}, vol.~25, no.~5, pp.
  392--410, Oct. 1977.

\bibitem{Wicker95}
S.~B. Wicker, \emph{Error Control Systems for Digital Communications and
  Storage}.\hskip 1em plus 0.5em minus 0.4em\relax Upper Saddle River, NJ:
  Prentice Hall, 1995.

\bibitem{Wang88}
Y.~Wang and X.~Zhu, ``A fast algorithm for the {F}ourier transform over finite
  fields and its {VLSI} implementation,'' \emph{{IEEE} J. Sel. Areas Commun.},
  vol.~6, no.~3, pp. 572--577, Apr. 1988.

\bibitem{Wicker94}
S.~B. Wicker and V.~K. Bhargava, Eds., \emph{Reed--Solomon Codes and Their
  Applications}.\hskip 1em plus 0.5em minus 0.4em\relax New York, NY: IEEE
  Press, 1994.

\bibitem{Brigham88}
E.~O. Brigham, \emph{The Fast Fourier Transform and Its Applications}.\hskip
  1em plus 0.5em minus 0.4em\relax Englewood, NJ: Prentice-Hall, 1988.

\bibitem{Sunar99}
B.~Sunar and {\c C}.~K. Ko{\c c}, ``Mastrovito multiplier for all trinomials,''
  \emph{{IEEE} Trans. Comput.}, vol.~48, no.~5, pp. 522--527, May 1999.

\bibitem{Mahboob05}
A.~Mahboob and N.~Ikram, ``Lookup table based multiplication technique for
  $\mathrm{GF}(2^m)$ with cryptographic significance,'' \emph{IEE
  Proc.-Commun.}, vol. 152, no.~6, pp. 965--974, Dec. 2005.

\end{thebibliography}
\end{document}